\documentclass[a4paper,11pt]{article}
\pdfoutput=1
\usepackage{jheppub}
\DeclareMathOperator{\Tr}{Tr}

\title{Anomalous dimension of the heavy-light quark current\\
in HQET up to four loops}
\author{Andrey Grozin}
\affiliation{Budker Institute of Nuclear Physics,\\
Lavrentiev St.~11, Novosibirsk 630090, Russia}
\emailAdd{A.G.Grozin@inp.nsk.su}
\abstract{The anomalous dimension of the heavy-light quark current in HQET is calculated up to four loops.
The N$^3$LL perturbative correction to $f_B/f_D$ is obtained.}

\begin{document} 
\maketitle
\flushbottom

\section{Introduction}
\label{S:intro}

Suppose we are interested in the quark current
\begin{equation}
j^{(5)}_0 = \bar{q}_0^{(5)} \Gamma b_0^{(5)} = Z_j^{(5)}(\alpha_s^{(5)}(\mu)) j^{(5)}(\mu)
\label{intro:j}
\end{equation}
in QCD$^{(5)}$ (QCD with $n_f=5$),
where $q$ is a light-quark field, and $\Gamma$ is a Dirac matrix.
For example, we want to obtain the matrix element $\langle0|j^{(5)}(\mu)|\bar{B}\rangle$
($p_{\bar{B}} = m_B v$).
Instead of the vacuum we can have, e.\,g., a light meson with a momentum $p \ll m_B$.
This problem is difficult because it contains a large scale $m_b$ plus several smaller scales.
We can eliminate this large scale by using Heavy Quark Effective Theory
(HQET, see, e.\,g., \cite{Neubert:1993mb,Manohar:2000dt,Grozin:2004yc}).
We do the following steps:

\begin{description}
\item[Running] Express $j^{(5)}(\mu)$ via $j^{(5)}(m_b)$ ($m_b$ is the on-shell mass).
The  vector currents $\bar{q} \gamma^\alpha b$ doesn't depend on the renormalization scale $\mu$:
$\gamma_j^{(5)}(\alpha_s^{(5)}) = d\log Z_j(\alpha_s^{(5)})/d\log\mu = 0$;
moreover, it is the same for all $n_f$, so that we can omit the upper index $(5)$.
The scalar current $(\bar{q} b)^{(n_f)}_\mu$ has $\gamma_j^{(n_f)} = -\gamma_m^{(n_f)}$;
the mass anomalous dimension $\gamma_m^{(n_f)}(\alpha_s^{(n_f)}) = d\log Z_m^{(n_f)}(\alpha_s^{(n_f)})/d\log\mu$
($m_0^{(n_f)} = Z_m^{(n_f)}(\alpha_s^{(n_f)}(\mu)) m^{(n_f)}(\mu)$)
is known up to five loops~\cite{Baikov:2014qja,Luthe:2016xec,Baikov:2017ujl}.
Multiplying $\Gamma$ by $\gamma_5^{\text{AC}}$ (the anticommuting $\gamma_5$) does not change the current's anomalous dimension.
We have $(\bar{q} \gamma_5^{\text{AC}} b)^{(n_f)}_\mu = Z_P^{(n_f)}(\alpha_s^{(n_f)}(\mu)) (\bar{q} \gamma_5^{\text{HV}} b)^{(n_f)}_\mu$,
$(\bar{q} \gamma_5^{\text{AC}} \gamma^\alpha b)^{(n_f)}_\mu = Z_A^{(n_f)}(\alpha_s^{(n_f)}(\mu)) (\bar{q} \gamma_5^{\text{HV}} \gamma^\alpha b)^{(n_f)}_\mu$,
$(\bar{q} \gamma_5^{\text{AC}} \sigma^{\alpha\beta} b)^{(n_f)}_\mu = (\bar{q} \gamma_5^{\text{HV}} \sigma^{\alpha\beta} b)^{(n_f)}_\mu$
(where $\gamma_5^{\text{HV}}$ is the 't~Hooft--Veltman $\gamma_5$);
the finite renormalization constants $Z^{(n_f)}_{P,A}(\alpha_s^{(n_f)})$ are known up to three loops~\cite{Larin:1991tj,Larin:1993tq}.
The renormalized currents $(\bar{q} \gamma_5^{\text{HV}} b)^{(n_f)}_\mu$, $(\bar{q} \gamma_5^{\text{HV}} \gamma^\alpha b)^{(n_f)}_\mu$
are related to $(\bar{q} \gamma^{[\alpha} \gamma^{\beta\vphantom{]}} \gamma^{\gamma\vphantom{[}} \gamma^{\delta]} b)_\mu^{(n_f)}$,
$(\bar{q} \gamma^{[\alpha} \gamma^{\beta\vphantom{]}} \gamma^{\gamma]} b)_\mu^{(n_f)}$
by the ordinary four-dimensional formulas with $\varepsilon_{\alpha\beta\gamma\delta}$
(square brackets mean antisymmetrization).
Hence the anomalous dimensions of the currents $j$
with $\Gamma = \gamma^{[\alpha} \gamma^{\beta\vphantom{]}} \gamma^{\gamma\vphantom{[}} \gamma^{\delta]}$
and $\Gamma = \gamma^{[\alpha} \gamma^{\beta\vphantom{]}} \gamma^{\gamma]}$
are known up to four loops.
Finally, the anomalous dimension of the tensor current ($\Gamma = \gamma^{[\alpha} \gamma^{\beta]}$)
is also known up to four loops~\cite{Baikov:2006ai,Gracey:2022vqr}.
In addition to the anomalous dimensions, one also needs the $\beta$ function
for solving the renormalization group equations;
it is known up to five loops~\cite{Herzog:2017ohr,Luthe:2017ttg,Chetyrkin:2017bjc}.
\item[Matching] Express $j^{(5)}(m_b)$ via HQET$^{(4)}$ operators:
\begin{align}
&j^{(5)}(m_b) = C_\Gamma^{(4)}(m_b) \tilde{\jmath}^{(4)}(m_b)
+ \frac{1}{2 m_b} \sum_i B_{\Gamma i}^{(4)}(m_b) O_i^{(4)}(m_b)
+ \mathcal{O}\biggl(\frac{1}{m_b^2}\biggr)\,,
\label{intro:match}\\
&\tilde{\jmath}_0^{(4)} = \bar{q}_0^{(4)} \Gamma \tilde{b}_0^{(4)} = \tilde{Z}_j^{(4)}(\alpha_s^{(4)}(\mu)) \tilde{\jmath}^{(4)}(\mu)\,,
\nonumber
\end{align}
where $\tilde{b}$ is the HQET static field with the velocity $v$
(we consider parts of $\Gamma$ commuting and anticommuting with $\rlap/v$ separately
in order to have a single leading-power term in~(\ref{intro:match})).
Here $O_i$ are the dimension four HQET$^{(4)}$ operators
with appropriate quantum numbers~\cite{Golden:1990dx,Neubert:1993za,Campanario:2003ix}.
The QCD/HQET matching coefficients $C_\Gamma^{(n_f)}$ are known at one~\cite{Eichten:1989zv},
two~\cite{Broadhurst:1994se,Grozin:1998kf}
and three~\cite{Bekavac:2009zc} loops.
\item[Running] Express $\tilde{\jmath}^{(4)}(m_b)$ via $\tilde{\jmath}^{(4)}(\mu)$
using $\tilde{\gamma}_j^{(4)}(\alpha_s^{(4)}) = d\log\tilde{Z}_j^{(4)}(\alpha_s^{(4)})/\log\mu$.
This anomalous dimension does not depend on $\Gamma$,
and is known at one~\cite{Voloshin:1986dir,Politzer:1988wp,Politzer:1988bs},
two~\cite{Ji:1991pr,Broadhurst:1991fz,Gimenez:1991bf}
and three~\cite{Chetyrkin:2003vi} loops.
We can stop running at some $\mu\in[m_c,m_b]$
and find $\langle0|\tilde{\jmath}^{(4)}(\mu)|\bar{B}\rangle$
using, e.\,g., lattice simulations or QCD sum rules.
But this problem still contains a large scale $m_c$.
So, we can run down to $m_c$ and then eliminate this scale.
\item[Matching] Express $\tilde{\jmath}^{(4)}(m_c)$ via HQET$^{(3)}$ operators:
\begin{equation}
\tilde{\jmath}^{(4)}(m_c) = \tilde{C}^{(3)}(m_c) \tilde{\jmath}^{(3)}(m_c)
+ \frac{1}{2 m_c} \sum_i \tilde{B}_i^{(3)}(m_c) O_i^{(3)}(m_c)
+ \mathcal{O}\biggl(\frac{1}{m_c^2}\biggr)\,,
\label{intro:match2}
\end{equation}
where $O_i^{(3)}$ are the dimension four HQET$^{(3)}$ operators
with appropriate quantum numbers.
The matching coefficient $\tilde{C}^{(3)}(m_c)$ (which does not depend on $\Gamma$)
is known at two~\cite{Grozin:1998kf} and three~\cite{Grozin:2006xm} loops.
The matching coefficients $\tilde{B}_i^{(3)}(m_c)$ (as well as $\tilde{C}^{(3)}(m_c)-1$)
come from diagrams with a $c$-quark loop, and hence start from two loops ($\alpha_s^2$).
\item[Running] Express $\tilde{\jmath}^{(3)}(m_c)$ via $\tilde{\jmath}^{(3)}(\mu)$ at some low $\mu$.
The anomalous dimension $\tilde{\gamma}_j^{(3)}(\alpha_s^{(3)})$ in HQET$^{(3)}$
is given by the same formula as in HQET$^{(4)}$, just $n_f$ differs.
Now we have only low scales $\mu$ and $\Lambda^{(3)}_{\overline{\text{MS}}}$,
and can use lattice simulations or QCD sum rules to find matrix elements.
\end{description}

So, all matching coefficients are known up to three loops.
For consistency, they should be used with the four-loop anomalous dimension (see section~\ref{S:ratio}).
In this article we obtain this previously unknown four-loop term.
As an application, we consider $f_B/f_D$ in section~\ref{S:ratio}.

\section{Calculation}
\label{S:calc}

In this section and the next one we'll live in a single HQET$^{(n_f)}$ theory
(with $n_f$ \emph{dynamic} flavors),
and hence we'll omit all upper indices $(n_f)$.
Due to superflavor symmetry~\cite{Georgi:1990ak},
we can use a spin $0$ static field $\tilde{Q}$ with velocity $v$.
We assume that all light quarks are massless ---
this assumption does not influence the anomalous dimension $\tilde{\gamma}_j$.
Let's define the vertex function $\tilde{\Gamma}(\omega,p)$ as the sum of all one-particle-irreducible diagrams
with an insertion of the current $\tilde{\jmath}_0 = \bar{q}_0 \tilde{Q}_0$,
the incoming HQET line with residual energy $\omega$
and the outgoing light-quark line with momentum $p$.
Each diagram has an even number of $\gamma$ matrices on the external fermion line,
and hence the only possible Dirac structures of $\tilde{\Gamma}(\omega,p)$
are $1$ and $[\rlap/v,\rlap/p]$.
For our purpose we can consider $\tilde{\Gamma}(\omega) \equiv \tilde{\Gamma}(\omega,0)$,
because $\omega < 0$ is sufficient to ensure infrared finiteness.
So, $\tilde{\Gamma}(\omega)$ is a scalar function;
it contains a single scale $\omega$.

We use dimensional regularization ($d=4-2\varepsilon$)
and $\overline{\text{MS}}$ renormalization.
Our aim is to obtain
$\tilde{Z}_j(\alpha_s) = \tilde{Z}_Q^{1/2}(\alpha_s,a) Z_q^{1/2}(\alpha_s,a) \tilde{Z}_\Gamma(\alpha_s,a)$,
where $\tilde{Z}_\Gamma(\alpha_s,a)$ is defined as following.
If we re-express $\tilde{\Gamma}(\omega)$ via the renormalized quantities $\alpha_s(\mu)$ and $a(\mu)$
instead of the bare quantities $g_0^2$ and $a_0$ (where $a_0 = Z_A(\alpha_s(\mu),a(\mu)) a(\mu)$
is the gauge parameter, $Z_A(\alpha_s,a)$ is the gluon field renormalization constant),
it becomes $\tilde{Z}_\Gamma(\alpha_s(\mu),a(\mu)) \tilde{\Gamma}(\omega;\mu)$,
where the renormalized vertex function $\tilde{\Gamma}(\omega;\mu)$ is finite at $\varepsilon\to0$.
In other words,
\begin{equation}
\log\tilde{\Gamma}(\omega) = \log\tilde{Z}_\Gamma(\alpha_s(\mu),a(\mu)) + \mathcal{O}(\varepsilon^0)\,.
\label{calc:ZV}
\end{equation}
Note that $\log\tilde{Z}_\Gamma(\alpha_s(\mu),a(\mu))$ must not depend on $\omega$,
while terms with non-negative powers of $\varepsilon$, of course, do depend on $\omega$.
So, the anomalous dimension is
\begin{equation}
\tilde{\gamma}_j(\alpha_s) = \tilde{\gamma}_\Gamma(\alpha_s,a)
+ \frac{1}{2} \bigl[\tilde{\gamma}_Q(\alpha_s,a) + \gamma_q(\alpha_s,a)\bigr]\,,
\label{calc:gamma}
\end{equation}
where $\tilde{\gamma}_\Gamma = d\log\tilde{Z}_\Gamma/d\log\mu$, $\tilde{\gamma}_Q = d\log\tilde{Z}_Q/d\log\mu$,
$\gamma_q = d\log Z_q/d\log\mu$.
Note that $\tilde{\gamma}_\Gamma$, $\tilde{\gamma}_Q$, $\gamma_q$ taken separately
are not gauge invariant (they depend on $a$);
however, $\tilde{\gamma}_j$ is gauge invariant,
because $\tilde{\jmath}$ is a colorless local operator,
so that all terms with $a$ must cancel in~(\ref{calc:gamma}).
The anomalous dimension $\tilde{\gamma}_Q$ is known up to four loops~\cite{Grozin:2022wse};
we need $\gamma_q$ up to four loops ($\xi^0$ and $\xi^1$ terms),
and these terms have been obtained in~\cite{Czakon:2004bu}
(see~\cite{Baikov:2014qja,Luthe:2016xec,Baikov:2017ujl,Chetyrkin:2017bjc} for five-loop results
and the four-loop result exact in $\xi$).

We use the \textsf{Mathematica} package \textsf{LiteRed}~\cite{Lee:2012cn,Lee:2013mka}
for reduction of diagrams to master integrals.
More exactly, we use its new version \textsf{LiteRed2} (\url{https://github.com/rnlg/LiteRed2});
new features of this version are crucial for the calculation.
A large portion of HQET diagrams for $\tilde{\Gamma}(\omega)$ contain linearly dependent denominators.
The package \textsf{LiteRed2} allows the user to define a \emph{set} of scalar Feynman integrals
with (possibly) dependent denominators.
Families of scalar integrals with linearly independent denominators are called \emph{bases}
(in physical literature they are often called generic topologies).
\textsf{LiteRed2} can find external symmetries of sectors of a \emph{set}
and sectors of several \emph{bases},
and provides the mappings of the integration momenta which implement these symmetries.
It also implements the A.~Pak's partial-fractioning algorithm~\cite{Pak:2011xt}.

After eliminating linearly dependent denominators,
there are 19 families (\emph{bases}) of scalar integrals.
Using integration by parts,
they can be reduced to 54 master integrals~\cite{Lee:2022art}.
Of these master integrals, 13 are recursively one-loop (hence simple combinations of $\Gamma$ functions);
10 can be expressed via ${}_3F_2$ hypergeometric functions of unit argument
using formulas from~\cite{Beneke:1994sw,Grozin:2000jv,Grozin:2003ak}
(in one case the ${}_3F_2$ happens to be expressible via $\Gamma$ functions, nobody knows why);
for 2 master integrals, a few terms of their $\varepsilon$ expansions
can be obtained from~\cite{Czarnecki:2001rh}.
Expansions of all 54 master integrals in $\varepsilon$ up to high orders
(up to weight 12 terms) have been obtained in~\cite{Lee:2022art},
using DRA method~\cite{Lee:2009dh}.

We use the variant of the QCD Feynman rules without the four-gluon vertex,
but with an auxiliary antisymmetric tensor field $t^a_{\mu\nu}$
whose propagator does not depend on its momentum;
it interacts with gluons via a $tAA$ vertex~\cite{Pukhov:1999gg}.
Then each diagram factorizes into a color factor and a loop integral;
its integrand consists of the Lorentz factors of all its vertices and propagators.

We use the covariant gauge:
the gluon propagator is $(1/k^2) (g_{\mu\nu} - \xi k_\mu k_\nu/k^2)$, $\xi=1-a_0$.
Up to three loops, we keep all powers of $\xi$;
at four loops, we keep only $\xi^0$ and $\xi^1$.
Higher powers of $\xi$ would produce many more terms in the loop integrands
and higher degrees of gluon denominators, thus making IBP reduction more difficult.
In principle, we could do the whole calculation in Feynman gauge $\xi=0$,
because the result $\tilde{\gamma}_j$ is gauge invariant.
But keeping $\xi^1$ terms provides a good check:
we keep $\xi^0$ and $\xi^1$ in all terms in~(\ref{calc:gamma})
and check that $\xi^1$ terms have canceled.

We use \textsf{qgraf}~\cite{Nogueira:1991ex} to generate all $L$-loop diagrams ($L\le4$)
for $\tilde{\Gamma}(\omega)$
(at four loops there are 7632 diagrams).
For each diagram \textsf{qgraf} produces three \textsf{form}~\cite{Kuipers:2012rf,Ruijl:2017dtg} sources.

\begin{itemize}
\item The first one contains the product of the color factors of all the vertices and propagators in the diagram.
Using the \textsf{form} package \textsf{color}~\cite{vanRitbergen:1998pn} we obtain the color factors of all diagrams
(at four loops 445 diagrams having zero color factors are eliminated).
\item The second one contains the product of the denominators of all the propagators
expressed via the loop momenta chosen by \textsf{qgraf}.
Diagrams having identical sets of denominators are combined to groups
(at four loops there are 3063 such groups).
For each group, we use \textsf{LiteRed2} to define the corresponding \emph{set};
\textsf{LiteRed2} also provides extra factors which can appear only in numerators
so that all scalar products of the vectors can be written as linear combinations
of the denominators and these numerator factors.
Groups of diagrams whose \emph{set} contains only trivial (zero) sectors are eliminated
(at four loops 1661 groups remain).
For each non-zero group the \textsf{Mathematica} program produces the ``multiplication table''
of the vectors (the loop momenta $k_{1,\ldots,4}$ and $v$) --- a list of substitutions replacing scalar products
of the vectors by linear combinations of the denominators (and the extra numerator factors).
Of course, when the denominators are linearly dependent, such expressions are not unique.
The program chooses one possible set of substitutions (using some systematic algorithm).
\textsf{LiteRed2} obtains mappings of all non-zero sectors of each \emph{set} to sectors
of the 19 families of scalar integrals with linearly-independent denominators.
\item The third one contains the product of the Lorentz structures of all the vertices and propagators in the diagram.
The \textsf{form} program finds HQET loops (if at least one is found, the diagram vanishes and is discarded);
finds all quark loops (and calculates the corresponding Dirac traces);
contracts all Lorentz indices, thus producing the integrand expressed via scalar products.
Using the corresponding ``multiplication table'' from the previous step,
the integrand is expressed via the denominators only.
\end{itemize}

Then \textsf{LiteRed2} transforms expressions for each diagram via scalar integrals belonging to 1661 \emph{sets}
(possibly with linear dependent denominators)
to combinations of scalar integrals belonging to 19 \emph{bases}
(with linearly independent denominators)
by partial fractioning.
All unique scalar integrals for each \emph{basis} are collected into a list,
and \textsf{LiteRed2} reduces them to the master integrals
(there are 183647 unique four-loop scalar integrals).
The global substitution list replacing all the scalar integrals by the corresponding linear combinations
of the master integrals is produced.
Each diagram is expressed via the master integrals.
Finally, $\tilde{\Gamma}(\omega)$ is calculated via the color factors and the master integrals,
and $\varepsilon$ expansions of the master integrals~\cite{Lee:2022art} are substituted.

In order to have a good check, we have calculated $\tilde{\gamma}_Q$ by the same set of programs.
Up to three loops, the result agrees with~\cite{Melnikov:2000zc,Chetyrkin:2003vi}.
At four loops, we obtain only $\xi^0$ and $\xi^1$ terms,
and they agree with the corresponding terms in~\cite{Grozin:2022wse}.

All pieces of the calculation are glued together by \emph{ad hoc} \textsf{python} scripts
orchestrated by a \textsf{Makefile}.
All calculations were done on a normal notebook, no supercomputer was used.
The total CPU time was about several days.

\section{Result}
\label{S:res}

The color factors are expressed via
\begin{align}
&\Tr t_R^a t_R^b = T_R \delta^{ab}\,,\quad
t_R^a t_R^a = C_R \mathbf{1}_R\,,\quad
N_R = \Tr \mathbf{1}_R\,,
\nonumber\\
&d_{RR'} = \frac{d_R^{abcd} d_{R'}^{abcd}}{N_R}\,,\quad
d_R^{abcd} = \Tr t_R^{(a} t_R^{b\vphantom{(}} t_R^{c\vphantom{)}} t_R^{d)}\,,
\label{res:color}
\end{align}
where $R = F$, $A$ are representations, and brackets mean symmetrization.
For $SU(N_c)$ gauge group with the standard normalization $T_F = \frac{1}{2}$ they are
\begin{align}
&C_F = \frac{N_c^2-1}{2N_c}\,,\quad
C_A = N_c\,,
\nonumber\\
&d_{FF} = \frac{(N_c^2-1) (N_c^4-6N_c^2+18)}{96 N_c^3}\,,\quad
d_{FA} = \frac{(N_c^2-1) (N_c^2+6)}{48}\,.
\label{res:SUN}
\end{align}

The anomalous dimension of the HQET heavy-light current is
\begin{align}
&\tilde{\gamma}_j(\alpha_s) = - 3 C_F \frac{\alpha_s}{4\pi}
+ C_F \biggl(\frac{\alpha_s}{4\pi}\biggr)^{\!2}
\biggl[ - C_F \biggl(\frac{8}{3} \pi^2 - \frac{5}{2}\biggr)
+ \frac{C_A}{3} \biggl(2 \pi^2 - \frac{49}{2}\biggr)
+ \frac{10}{3} T_F n_f\biggr]
\nonumber\\
&{} + C_F \biggl(\frac{\alpha_s}{4\pi}\biggr)^{\!3}
\biggl[ - C_F^2 \biggl(36 \zeta_3 + \frac{8}{9} \pi^4 - \frac{32}{3} \pi^2 + \frac{37}{2}\biggr)
\nonumber\\
&{} + \frac{C_F C_A}{3} \biggl(142 \zeta_3 - \frac{8}{15} \pi^4 - \frac{592}{9} \pi^2 - \frac{655}{12}\biggr)
- \frac{C_A^2}{3} \biggl(22 \zeta_3 + \frac{4}{5} \pi^4 - \frac{130}{9} \pi^2 - \frac{1451}{36}\biggr)
\nonumber\\
&{} - \frac{2}{3} C_F T_F n_f \biggl(88 \zeta_3 - \frac{112}{9} \pi^2 - \frac{235}{3}\biggr)
+ \frac{8}{3} C_A T_F n_f \biggl(19 \zeta_3 - \frac{7}{9} \pi^2 - \frac{64}{9}\biggr)
+ \frac{140}{27} (T_F n_f)^2\biggr]
\nonumber\\
&{} + \biggl(\frac{\alpha_s}{4\pi}\biggr)^{\!4}
\biggl[C_F^4 \biggl(1200 \zeta_5 - 168 \zeta_3^2 - \frac{896}{3} \pi^2 \zeta_3 + 394 \zeta_3
  + \frac{3884}{2835} \pi^6 - \frac{4}{15} \pi^4 + \frac{136}{3} \pi^2 - \frac{691}{8}\biggr)
\nonumber\\
&{} - C_F^3 C_A \biggl(\frac{5660}{3} \zeta_5 - 192 \zeta_3^2 - \frac{4576}{9} \pi^2 \zeta_3 + 1275 \zeta_3
  + \frac{2659}{2835} \pi^6 - \frac{119}{45} \pi^4 + \frac{2398}{9} \pi^2 - \frac{3991}{12}\biggr)
\nonumber\\
&{} + C_F^2 C_A^2 \biggl(\frac{434}{3} \zeta_5 - 42 \zeta_3^2 - \frac{1916}{9} \pi^2 \zeta_3 + \frac{39047}{27} \zeta_3
  + \frac{2087}{1890} \pi^6 - \frac{2663}{90} \pi^4 + \frac{41026}{243} \pi^2 - \frac{189671}{324}\biggr)
\nonumber\\
&{} + C_F C_A^3 \biggl(492 \zeta_5 + 30 \zeta_3^2 + \frac{352}{9} \pi^2 \zeta_3 - \frac{14666}{27} \zeta_3
  - \frac{1439}{8505} \pi^6 + \frac{23}{90} \pi^4 - \frac{7246}{243} \pi^2 + \frac{179089}{648}\biggr)
\nonumber\\
&{} + 8 d_{FA} \biggl(30 \zeta_5 + \frac{106}{3} \pi^2 \zeta_3 - 16 \zeta_3
  - \frac{452}{567} \pi^6 + \frac{29}{9} \pi^4 + \frac{46}{3} \pi^2 - 8\biggr)
\nonumber\\
&{} + 4 C_F^3 T_F n_f \biggl(\frac{580}{3} \zeta_5 - \frac{224}{9} \pi^2 \zeta_3 - 24 \zeta_3
  - \frac{29}{45} \pi^4 + \frac{68}{3} \pi^2 - \frac{119}{3}\biggr)
\nonumber\\
&{} - \frac{C_F^2 C_A T_F n_f}{3} \biggl(1096 \zeta_5 - \frac{736}{3} \pi^2 \zeta_3 + \frac{18980}{9} \zeta_3
  - \frac{1138}{45} \pi^4 - \frac{9404}{81} \pi^2 - \frac{32093}{27}\biggr)
\nonumber\\
&{} - C_F C_A^2 T_F n_f \biggl(308 \zeta_5 + 24 \zeta_3^2 + \frac{128}{9} \pi^2 \zeta_3 - \frac{20792}{27} \zeta_3
  - \frac{874}{8505} \pi^6 + \frac{56}{27} \pi^4 + \frac{5240}{243} \pi^2 + \frac{27269}{162}\biggr)
\nonumber\\
&{} - 32 d_{FF} n_f \biggl(15 \zeta_5 + \frac{8}{3} \pi^2 \zeta_3 - 8 \zeta_3
  - \frac{437}{2835} \pi^6 + \frac{4}{9} \pi^4 + \frac{20}{3} \pi^2 - 4\biggr)
\nonumber\\
&{} + \frac{16}{27} C_F^2 (T_F n_f)^2 \biggl(326 \zeta_3 - \frac{11}{5} \pi^4 + \frac{16}{9} \pi^2 - \frac{206}{3}\biggl)
\nonumber\\
&{} - \frac{2}{27} C_F C_A (T_F n_f)^2 \biggl(2272 \zeta_3 - \frac{76}{5} \pi^4 + \frac{32}{9} \pi^2 - \frac{761}{3}\biggr)
\nonumber\\
&{} - \frac{8}{9} C_F (T_F n_f)^3 \biggl(16 \zeta_3 - \frac{83}{9}\biggr)\biggr]
+ \mathcal{O}(\alpha_s^5)
\label{res:gamma}
\end{align}
Up to three loops, it agrees with~\cite{Chetyrkin:2003vi}.
Terms with the highest degrees of $n_f$ are known to all orders in $\alpha_s$~\cite{Broadhurst:1994se};
the $C_F (T_F n_f)^3 \alpha_s^4$ term in~(\ref{res:gamma}) agrees with this result.
All the remaining four-loop terms are new.
The highest weight at $L$ loops is $2(L-1)$, at least up to $L=4$.

The anomalous dimension~(\ref{res:gamma}) in the \textsf{Mathematica} syntax
is attached to this article as the file \textsf{gammaj.m}.
The notations used are explained in comments at the top of this file.

For the physical $SU(3)$ color group, the numerical result is
\begin{align}
&\tilde{\gamma}_j = - \frac{\alpha_s}{\pi}
+ (0.138889 n_f - 3.043282) \biggl(\frac{\alpha_s}{\pi}\biggr)^{\!2}
\nonumber\\
&{} + (0.027006 n_f^2 + 1.554061 n_f - 12.941040) \biggl(\frac{\alpha_s}{\pi}\biggr)^{\!3}
\nonumber\\
&{} + (- 0.005793 n_f^3 - 0.168484 n_f^2 + 12.158867 n_f - 59.446998) \biggl(\frac{\alpha_s}{\pi}\biggr)^{\!4}\,.
\label{res:num}
\end{align}
At $n_f=4$,
\begin{equation}
\tilde{\gamma}_j = - \frac{\alpha_s}{\pi}
- 2.487726 \biggl(\frac{\alpha_s}{\pi}\biggr)^{\!2}
- 6.292698 \biggl(\frac{\alpha_s}{\pi}\biggr)^{\!3}
- 13.878042 \biggl(\frac{\alpha_s}{\pi}\biggr)^{\!4}\,.
\label{res:num4}
\end{equation}

At the leading large $\beta_0$ order
($b = \beta_0 \alpha_s/(4\pi) \sim 1$, $1/\beta_0 \ll 1$, see, e.\,g., chapter~8 in~\cite{Grozin:2004yc})
we have~\cite{Broadhurst:1994se}
\begin{align}
&\tilde{\gamma}_j = - C_F \frac{b}{\beta_0}
\frac{\bigl(1+\frac{2}{3}b\bigr) \Gamma(4+2b)}{\Gamma^2(2+b) \Gamma(3+b) \Gamma(1-b)}
+ \mathcal{O}\biggl(\frac{1}{\beta_0^2}\biggr)
\label{res:lb0}\\
&{} - 3 C_F \frac{b}{\beta_0} \biggl[1 + \frac{5}{6} b - \frac{35}{36} b^2
- \biggl(2 \zeta_3 - \frac{83}{72}\biggr) b^3
- \biggl(5 \zeta_3 - \frac{\pi^4}{10} + \frac{65}{16}\biggr) \frac{b^4}{3}
+ \mathcal{O}(b^5)\biggr] + \mathcal{O}\biggl(\frac{1}{\beta_0^2}\biggr)\,.
\nonumber
\end{align}
At $n_f=4$,
\begin{equation}
\tilde{\gamma}_j = - \frac{\alpha_s}{\pi}
- 1.736111 \biggl(\frac{\alpha_s}{\pi}\biggr)^{\!2}
+ 4.219715 \biggl(\frac{\alpha_s}{\pi}\biggr)^{\!3}
+ 11.314887 \biggl(\frac{\alpha_s}{\pi}\biggr)^{\!4}
+ 2.083958 \biggl(\frac{\alpha_s}{\pi}\biggr)^{\!5}
+ \cdots\,.
\label{res:lb04}
\end{equation}
This approximation usually works rather well for matching coefficients
and other renormalized matrix elements (which usually contain renormalons),
but absolutely does not work for anomalous dimensions.

\section{The ratio $f_B/f_D$}
\label{S:ratio}

The decay constant $f_B$ is defined by $\langle0|j^{(5)}|\bar{B}\rangle = m_B f_B$,
where the current $j$ has Dirac structure $\Gamma = \gamma_5^{\text{AC}} \rlap/v$.
In HQET we need to use non-relativistic normalization of states
$|\bar{B}\rangle = \sqrt{2 m_B} |\bar{B}\rangle_{\text{nr}}$:
$\langle0|\tilde{\jmath}^{(4)}(\mu)|\bar{B}\rangle_{\text{nr}} = F^{(4)}(\mu)$, and
\begin{align}
f_B = {}& \sqrt{\frac{2}{m_B}} C_{\rlap/v}^{(4)}(m_b) F^{(4)}(m_b)
\nonumber\\
&{} \times \biggl[1 + \frac{1}{2 m_b}
\Bigl(C_{\rlap/v,\Lambda}^{(4)}(m_b) \bar{\Lambda} + G_k^{(4)}(m_b) + C_m^{(4)}(m_b) G_m^{(4)}(m_b)\Bigr)
+ \mathcal{O}\biggl(\frac{1}{m_b^2}\biggr)\biggr]\,,
\label{ratio:fB}
\end{align}
where $\bar{\Lambda} = m_B-m_b$,
\begin{equation*}
\langle0|O_{j,k}^{(4)}(\mu)|\bar{B}\rangle_{\text{nr}} = F^{(4)}(\mu) G_k^{(4)}(\mu)\,,\quad
O_{j,k0}^{(4)} = \int dx\,T\{\tilde{\jmath}_0^{(4)}(0),O_{k0}^{(4)}(x)\}\,,
\end{equation*}
$G_m^{(4)}(\mu)$ is defined similarly,
$O_{k,m}$ are the kinetic energy operator and the chromomagnetic interaction operator
in the HQET Lagrangian
\begin{equation*}
L = \bar{\tilde{Q}}_0 i D \cdot v \tilde{Q}_0
+ \frac{O_{k0} + C_{m0} O_{m0}}{2 m_Q}
+ \mathcal{O}\biggl(\frac{1}{m_Q^2}\biggr)
\end{equation*}
(see, e.\,g., (3.17) in~\cite{Campanario:2003ix}).
The formula for $f_D$ is similar.
Running of $F^{(n_f)}(\mu)$ is given by the solution of the renormalization group equation:
\begin{align}
&F^{(n_f)}(\mu) = \hat{F}{}^{(n_f)} \biggl(\frac{\alpha_s^{(n_f)}(\mu)}{4\pi}\biggr)^{\!\tilde{\gamma}_{j0}/(2\beta_0^{(n_f)})}
K^{(n_f)}(\alpha_s^{(n_f)}(\mu))\,,
\label{ratio:RG}\\
&K^{(n_f)}(\alpha_s) = \exp \int_0^{\alpha_s} \frac{d\alpha_s}{\alpha_s}
\biggl(\frac{\tilde{\gamma}_j^{(n_f)}(\alpha_s)}{2 \beta^{(n_f)}(\alpha_s)}
- \frac{\tilde{\gamma}_{j0}}{2 \beta_0^{(n_f)}}\biggr)\,.
\nonumber
\end{align}
Here
\begin{align*}
&\beta^{(n_f)}(\alpha_s^{(n_f)}) = - \frac{1}{2} \frac{d\log\alpha_s^{(n_f)}}{d\log\mu}
= \sum_{L=1}^\infty \beta_{L-1}^{(n_f)} \biggl(\frac{\alpha_s^{(n_f)}}{4\pi}\biggr)^{\!L}\,,\quad
\beta_0^{(n_f)} = \frac{11}{3} C_A - \frac{4}{3} T_F n_f\,,\\
&\tilde{\gamma}_j^{(n_f)}(\alpha_s) = \tilde{\gamma}_{j0} \frac{\alpha_s}{4\pi}
+ \sum_{L=2}^\infty \tilde{\gamma}_{j,L-1}^{(n_f)} \biggl(\frac{\alpha_s}{4\pi}\biggr)^{\!L}\,.
\end{align*}
So, the ratio $f_B/f_D$ is
\begin{align}
\frac{f_B}{f_D} = {}& \sqrt{\frac{m_D}{m_B}} \frac{C_{\rlap/v}^{(4)}(m_b)}{C_{\rlap/v}^{(3)}(m_c)} \tilde{C}{}^{(3)}(m_c)
\biggl(\frac{\alpha_s^{(4)}(m_b)}{\alpha_s^{(4)}(m_c)}\biggr)^{\!\tilde{\gamma}_{j0}/(2\beta_0^{(4)})}
\frac{K^{(4)}(\alpha_s^{(4)}(m_b))}{K^{(4)}(\alpha_s^{(4)}(m_c))}
\nonumber\\
&{} \times \biggl[1 + A \biggl(\frac{1}{m_c} - \frac{1}{m_b}\biggr) + \mathcal{O}\biggl(\frac{1}{m_{c,b}^2}\biggr)\biggr]\,.
\label{ratio:ratio}
\end{align}
The nonperturbative parameters $G_{k,m}$ were estimated from HQET sum rules~\cite{Neubert:1992fk,Ball:1994uh},
their precision is not high.
Therefore we use the tree-level values $C_{\rlap/v,\Lambda} = -1$, $C_m = 1$,
neglect running of $G_{k,m}$ and their differences between $n_f=4$ and $3$,
and neglect the $\alpha_s^2/m_c$ corrections in~(\ref{intro:match2}):
\begin{equation}
A = \frac{1}{2} \bigl(\bar{\Lambda} - G_k - G_m\bigr)\,.
\label{ratio:A}
\end{equation}

We obtain
\begin{align}
&\frac{f_B}{f_D} = \sqrt{\frac{m_D}{m_B}} x^{-\tilde{\gamma}_{j0}/(2\beta_0^{(4)})}
\biggl\{1 + r_1 (x-1) a_s + \biggl[r_{20} + r_{21} (x^2-1) + \frac{r_1^2}{2} (x-1)^2\biggr] a_s^2
\nonumber\\
&{} + \biggl[r_{30} + r_{31} (x^3-1) + \frac{r_1^3}{6} (x-1)^3 + r_1 r_{20} (x-1) + r_1 r_{21} (x-1) (x^2-1)\biggr] a_s^3
\nonumber\\
&{} + A \biggl(\frac{1}{m_c} - \frac{1}{m_b}\biggr)
+ \mathcal{O}\biggl(\alpha_s^4,\frac{1}{m_{c,b}^2}\biggr)\biggr\}\,,
\label{ratio:res}
\end{align}
where $a_s = \alpha_s^{(4)}(m_b)/(4\pi)$, $x = \alpha_s^{(4)}(m_c)/\alpha_s^{(4)}(m_b)$,
\begin{align*}
&r_1 = - c_1 - \frac{\tilde{\gamma}_{j0}}{2\beta_0^{(4)}}
\biggl(\frac{\tilde{\gamma}_{j1}^{(4)}}{\tilde{\gamma}_{j0}}
- \frac{\beta_1^{(4)}}{\beta_0^{(4)}}\biggr)\,,\quad
r_{20} = c_2^{(4)} - c_2^{(3)} + z_2\,,\\
&r_{21} = - c_2^{(3)} + \frac{c_1^2}{2} + z_2
+ \frac{\tilde{\gamma}_{j0}}{4\beta_0^{(4)}}
\biggl[ - \frac{\tilde{\gamma}_{j2}^{(4)}}{\tilde{\gamma}_{j0}}
+ \frac{\beta_1^{(4)}}{\beta_0^{(4)}} \frac{\tilde{\gamma}_{j1}^{(4)}}{\tilde{\gamma}_{j0}}
+ \frac{\beta_2^{(4)}}{\beta_0^{(4)}}
- \biggl(\frac{\beta_1^{(4)}}{\beta_0^{(4)}}\biggr)^{\!2}\biggr]\,,\\
&r_{30} = c_3^{(4)} - c_3^{(3)} - c_1 \bigl(c_2^{(4)} - c_2^{(3)} + d_2\bigr) + z_3\,,\\
&r_{31} = - c_3^{(3)} + c_1 (c_2^{(3)} - d_2) - \frac{c_1^3}{3} + z_3\\
&{} + \frac{\tilde{\gamma}_{j0}}{6 \beta_0^{(4)}}
\biggl[ - \frac{\tilde{\gamma}_{j3}^{(4)}}{\tilde{\gamma}_{j0}}
+ \frac{\beta_1^{(4)}}{\beta_0^{(4)}} \frac{\tilde{\gamma}_{j2}^{(4)}}{\tilde{\gamma}_{j0}}
+ \frac{\beta_2^{(4)}}{\beta_0^{(4)}} \frac{\tilde{\gamma}_{j1}^{(4)}}{\tilde{\gamma}_{j0}}
- \biggl(\frac{\beta_1^{(4)}}{\beta_0^{(4)}}\biggr)^{\!2} \frac{\tilde{\gamma}_{j1}^{(4)}}{\tilde{\gamma}_{j0}}
+ \frac{\beta_3^{(4)}}{\beta_0^{(4)}}
- 2 \frac{\beta_1^{(4)}}{\beta_0^{(4)}} \frac{\beta_2^{(4)}}{\beta_0^{(4)}}
+ \biggl(\frac{\beta_1^{(4)}}{\beta_0^{(4)}}\biggr)^{\!3}\biggr]\,,\\
&C_{\rlap/v}^{(n_f)}(m_Q) = 1 + c_1 \frac{\alpha_s^{(n_f)}(m_Q)}{4\pi}
+ \sum_{L=2}^\infty c_L^{(n_f)} \biggl(\frac{\alpha_s^{(n_f)}(m_Q)}{4\pi}\biggr)^{\!L}\,,\\
&\tilde{C}{}^{(n_f)}(m_Q) = 1 + z_2 \biggl(\frac{\alpha_s^{(n_f+1)}(m_Q)}{4\pi}\biggr)^{\!2}
+ \sum_{L=3}^\infty z_L^{(n_f)} \biggl(\frac{\alpha_s^{(n_f+1)}(m_Q)}{4\pi}\biggr)^{\!L}\,,\\
&\frac{\alpha_s^{(3)}(m_c)}{4\pi} = \frac{\alpha_s^{(4)}(m_c)}{4\pi}
\biggl[1 + \sum_{L=2}^\infty d_L \biggl(\frac{\alpha_s^{(4)}(m_c)}{4\pi}\biggr)^{\!L}\biggr]
\end{align*}
(where $m_Q = m_b$ for $n_f=4$ and $m_c$ for $n_f=3$),
cf.~(43--44) in~\cite{Grozin:2007fh}.
The terms up to $a_s^2$ were obtained in~\cite{Chetyrkin:2003vi},
the $a_s^3$ term is new.
The result~\cite{Bekavac:2009zc} (\url{https://www.ttp.kit.edu/Progdata/ttp09/ttp09-41/}) for $C_{\rlap/v}^{(n_f)}(m_Q)$
is expressed via $\alpha_s^{(n_f)}(m_Q)$,
and the result~\cite{Grozin:2006xm} (\url{https://www.ttp.kit.edu/Progdata/ttp06/ttp06-25/}) for $\tilde{C}^{(n_f)}(m_Q)$
via $\alpha_s^{(n_f+1)}(m_Q)$.
It would be more logical to express both results either via $\alpha_s^{(n_f)}(m_Q)$ or via $\alpha_s^{(n_f+1)}(m_Q)$.
But up to our accuracy level we may simply replace $\alpha_s^{(n_f+1)}(m_Q) \to \alpha_s^{(n_f)}(m_Q)$
in the formula for $\tilde{C}^{(n_f)}(m_Q)$~\cite{Grozin:2006xm}.
The coefficients $\beta_{L-1}^{(n_f)}$ for $L\le4$ have been obtained in~\cite{vanRitbergen:1997va,Czakon:2004bu}
(see~\cite{Herzog:2017ohr,Luthe:2017ttg,Chetyrkin:2017bjc} for the five-loop result);
the two-loop decoupling coefficient $d_2 = \bigl(- 15 C_F + \frac{32}{3} C_A\bigr) T_F$ is from~\cite{Chetyrkin:1997un}.
Terms in $C_\Gamma^{(4)}$ with $c$-quark loops are non-trivial functions of $m_c/m_b$;
for them, at two loops we use the exact formula~\cite{Broadhurst:1994se},
and at four loops --- the expansion up to $(m_c/m_b)^8$~\cite{Bekavac:2009zc}.
For the coupling constants we used \textsf{RunDec}~\cite{Chetyrkin:2000yt,Herren:2017osy} version 3.1
and got $\alpha_s^{(4)}(m_b) = 0.215$, $x = 1.63$.

Numerically (the sum rules result~\cite{Ball:1994uh} is $A \sim 1\,\text{GeV}$ with large errors)
\begin{align}
\frac{f_B}{f_D} = {}& 0.669 \cdot
\biggl[1 + 0.566 \frac{\alpha_s^{(4)}(m_b)}{\pi} + 6.176 \biggl(\frac{\alpha_s^{(4)}(m_b)}{\pi}\biggr)^{\!2}
+ 99.170 \biggl(\frac{\alpha_s^{(4)}(m_b)}{\pi}\biggr)^{\!3}
\nonumber\\
&{} + [\sim 1\,\text{GeV}]\cdot\biggl(\frac{1}{m_c} - \frac{1}{m_b}\biggr)\biggr]
\nonumber\\
= {}& 0.669 \cdot (1 + 0.039 + 0.029 + 0.032 + [\sim 0.46])
\label{ratio:num}
\end{align}
(an estimate of the first $1/m_{c,b}$ correction is in square brackets).
Convergence of the perturbative series is questionable,
though each perturbative correction is small.
If we omit the power correction, the result is $f_B/f_D = 0.669\cdot1.100 = 0.736$;
with the estimate of the power correction included, it is $0.669\cdot1.56 = 1.04$.

At the leading large $\beta_0$ order we have~\cite{Broadhurst:1994se} (see also chapter~8 in~\cite{Grozin:2004yc})
\begin{align}
&K(\alpha_s(m_b)) C_{\rlap/v}(m_b) = 1
+ \frac{1}{\beta_0} \int_0^\infty du\,e^{-u/b} S(u)
+ \mathcal{O}\biggl(\frac{1}{\beta_0^2}\biggr)\,,
\label{ratio:KC}\\
&S(u) = - 3 C_F \biggl[e^{\frac{5}{3}u} \frac{\Gamma(u) \Gamma(1-2u)}{\Gamma(3-u)} (1-u-u^2) - \frac{1}{2u}\biggr]\,,
\nonumber
\end{align}
where $b = \beta_0 a_s$.
At this accuracy level differences of various quantities for $n_f=4$ and $3$ can be neglected,
and $\tilde{C}(m_c) = 1$.
We obtain
\begin{align}
\frac{f_B}{f_D} ={}& \sqrt{\frac{m_D}{m_B}} x^{-\tilde{\gamma}_{j0}/(2\beta_0)}
\biggl[1 + \frac{1}{\beta_0} \int_0^\infty du \bigl(e^{-u/b} - e^{-u/(xb)}\bigr) S(u)
+ \mathcal{O}\biggl(\frac{1}{\beta_0^2}\biggr)\biggr]
\nonumber\\
&{} \times \biggl[1 + A \biggl(\frac{1}{m_c} - \frac{1}{m_b}\biggr) + \mathcal{O}\biggl(\frac{1}{m_{c,b}^2}\biggr)\biggr]\,.
\label{ratio:lb0}
\end{align}
Expanding $S(u)$ in $u$ we find the first square bracket in~(\ref{ratio:lb0}) as
\begin{align}
&1 + C_F \frac{b}{\beta_0} \biggl[\frac{13}{4} (x-1)
+ \frac{1}{2} \biggl(\pi^2 + \frac{53}{12}\biggr) (x^2-1) b
+ \biggl(6 \zeta_3 + \frac{13}{6} \pi^2 - \frac{751}{216}\biggr) (x^3-1) b^2
\nonumber\\
&{} + \biggl(39 \zeta_3 + \frac{9}{10} \pi^4 + \frac{53}{12} \pi^2 - \frac{16771}{432}\biggr) (x^4-1) b^3
+ \mathcal{O}(b^4)\biggr]
+ \mathcal{O}\biggl(\frac{1}{\beta_0^2}\biggr)\,.
\label{ratio:exp}
\end{align}
This expression reproduces all terms with the largest degrees of $n_f$ at each order in $\alpha_s$ in~(\ref{ratio:res}).
Numerically, (\ref{ratio:exp}) gives
\begin{align}
&1
+ 0.686 \frac{\alpha_s^{(4)}(m_b)}{\pi}
+ 8.271 \biggl(\frac{\alpha_s^{(4)}(m_b)}{\pi}\biggr)^{\!2}
+ 121.97 \biggl(\frac{\alpha_s^{(4)}(m_b)}{\pi}\biggr)^{\!3}
+ 2567.6 \biggl(\frac{\alpha_s^{(4)}(m_b)}{\pi}\biggr)^{\!4}
+ \cdots
\nonumber\\
&{} = 1 + 0.047 + 0.039 + 0.039 + 0.056 + \cdots
\label{ratio:expnum}
\end{align}
Comparing this series with~(\ref{ratio:num}) we see
that naive nonabelianization~\cite{Broadhurst:1994se} works rather well up to the N$^3$LL level.

Of course, the integral~(\ref{ratio:KC}) is ill defined due to IR renormalon poles at positive $u$.
We can use, e.\,g., the principal value prescription.
Other prescriptions would produce different results;
the residue at the leading renormalon pole $u=\frac{1}{2}$
is a measure of theoretical uncertainty~\cite{Broadhurst:1994se}:
\begin{equation*}
\Delta C_{\rlap/v}(\mu) = \frac{1}{4} \frac{\Delta\bar{\Lambda}}{m_Q}\,,
\text{ where }
\Delta\bar{\Lambda} = - 2 C_F \frac{e^{5/6} \Lambda_{\overline{\text{MS}}}}{\beta_0}
\end{equation*}
is the leading UV renormalon ambiguity of $\bar{\Lambda}$.
Using also the UV renormalon ambiguities
$\Delta G_k = - \frac{3}{2} \Delta\bar{\Lambda}$, $\Delta G_m = 2\,\Delta\bar{\Lambda}$~\cite{Neubert:1994wq}
(see also~\cite{Broadhurst:1994se} and chapter~8 in~\cite{Grozin:2004yc})
we see that the renormalon ambiguities in $f_B/f_D$ cancel~\cite{Neubert:1994wq}.
Each prescription for summing the divergent series~(\ref{ratio:KC})
corresponds to some values of $\bar{\Lambda}$, $G_k$, $G_m$;
when we change the prescription, these values change accordingly.
The sum of the divergent perturbative series~(\ref{ratio:expnum})
(the first square bracket in~(\ref{ratio:lb0}))
is, according to the principal value prescription, $1.077 \pm 0.025$.
Here the theoretical uncertainty is
\begin{equation*}
\frac{C_F}{2} \frac{e^{5/6} \Lambda_{\overline{\text{MS}}}^{(4)}}{\beta_0^{(4)}}
\biggl(\frac{1}{m_c} - \frac{1}{m_b}\biggr)\,,
\end{equation*}
and $\Lambda_{\overline{\text{MS}}}^{(4)} = 292\,\text{MeV}$, according to \textsf{RunDec} 3.1.
In other words, the all-orders leading large $\beta_0$ result (without power corrections)
is $f_B/f_D = 0.721 \pm 0.016$.

We can also try to estimate the sum of the divergent perturbative series~(\ref{ratio:num})
without resorting to the large $\beta_0$ limit.
This series can be written as
\begin{equation}
1 + c \alpha_s \biggl(1 + \sum_{n=1}^\infty c_n \alpha_s^n\biggr)
= 1 + c \int_0^\infty du\,e^{-u/\alpha_s} S(u)\,,\quad
S(u) = 1 + \sum_{n=1}^\infty c_n \frac{u^n}{n!}
\label{ratio:series}
\end{equation}
($\alpha_s \equiv \alpha_s^{(4)}(m_b)$).
Then we replace the series $S(u)$ by the Pad\'e approximant $(1 + p_1 u)/(1 + p_2 u)$,
where $p_{1,2}$ are obtained from the known coefficients $c_{1,2}$:
$S(u) = (1 + 0.917 u)/(1 - 2.555 u)$.
This rational function has a pole at $u_0 = 0.391$;
the radius of convergence of the series $S(u)$ is $u_0$.
Expanding the approximant we get
\begin{equation*}
1
+ 0.566 \frac{\alpha_s}{\pi}
+ 6.176 \biggl(\frac{\alpha_s}{\pi}\biggr)^{\!2}
+ 99.17 \biggl(\frac{\alpha_s}{\pi}\biggr)^{\!3}
+ 2388 \biggl(\frac{\alpha_s}{\pi}\biggr)^{\!4}
+ 76699 \biggl(\frac{\alpha_s}{\pi}\biggr)^{\!5}
+ \cdots
\end{equation*}
(the first 3 corrections coincide with~(\ref{ratio:num}) by construction,
the next ones are an extrapolation due to the Pad\'e approximant).
All corrections are positive, the coefficients grow fast.
We define the sum of this series as the principal value of the integral in $u$.
The estimate of the theoretical uncertainty is given by the residue at the pole $u = u_0$.
The sum of the perturbative series, estimated using this method, is $1.053 \pm 0.016$.
In other words, the all-orders result (without power corrections)
is $f_B/f_D = 0.705 \pm 0.010$.

The effect of the (poorly known) $1/m_{c,b}$ correction is large.
It would be interesting to extract $G_{k,m}$ from HQET lattice simulations.
The $1/m_{c,b}^2$ corrections (see~\cite{Falk:1992wt,Balzereit:1996cu}) also can be substantial
and deserve further investigation.

The lattice results~\cite{FlavourLatticeAveragingGroupFLAG:2021npn}
$f_B = (190.0 \pm 1.3)$\,MeV and $f_D = (212.0 \pm 0.7)$\,MeV
lead to $f_B/f_D = 0.896 \pm 0.009$
(the errors of $f_B$ and $f_D$ may be correlated, so, we have added the relative errors linearly;
if we believe that they are uncorrelated, the errors can be added quadratically, producing $\pm 0.007$).

\section{Conclusion}
\label{S:conc}

The anomalous dimension of the heavy-light quark current in HQET is now known up to four loops~(\ref{res:gamma}).
The perturbative corrections to the ratio $f_B/f_D$ are now known up to the N$^3$LL level~(\ref{ratio:res}). 

\acknowledgments

I am grateful to R.\,N.~Lee for numerous discussions
and consultations on \textsf{LiteRed2}.
The work has been supported by the Russian Science Foundation, grant number 20-12-00205.

\bibliographystyle{JHEP}
\bibliography{adim4.bib}

\end{document}